\documentclass[%
preprint,
superscriptaddress,
 amsmath,amssymb,
prl,
]{revtex4-1}

\usepackage{graphicx}
\usepackage{dcolumn}
\usepackage{bm}
\usepackage{lineno}
\usepackage{braket} 
\usepackage{siunitx} 
\usepackage{color}

\begin{document}


\title{Monitoring spin coherence of single nitrogen-vacancy centers in nanodiamonds during pH changes in aqueous buffer solutions}

\author{Masazumi Fujiwara}
 \address{Department of Chemistry, Osaka City University, Sumiyoshi-ku, Osaka 558-8585, Japan}
 \address{School of Science and Technology, Kwansei Gakuin University, Sanda Hyogo 669-1337, Japan}
  \email{masazumi@osaka-cu.ac.jp}
\author{Ryuta Tsukahara} 
 \affiliation{School of Science and Technology, Kwansei Gakuin University, Sanda Hyogo 669-1337, Japan}
\author{Yoshihiko Sera}
 \affiliation{School of Science and Technology, Kwansei Gakuin University, Sanda Hyogo 669-1337, Japan}
\author{Hiroshi Yukawa}
 \affiliation{Department of Biomolecular Engineering, Graduate School of Engineering, Nagoya University, Chikusa-ku, Nagoya 464-8603, Japan}
 \affiliation{Institute of Nano-Life-Systems, Institutes of Innovation for Future Society, Nagoya University, Chikusa-ku, Nagoya 464-8603, Japan}
\author{Yoshinobu Baba}
 \affiliation{Department of Biomolecular Engineering, Graduate School of Engineering, Nagoya University, Chikusa-ku, Nagoya 464-8603, Japan}
 \affiliation{Health Research Institute, National Institute of Advanced Industrial Science and Technology (AIST), Takamatsu 761-0395, Japan}
 \affiliation{College of Pharmacy, Kaohsiung Medical University, Kaohsiung, 807, Taiwan, R.O.C.}
\author{Shinichi Shikata}
 \affiliation{School of Science and Technology, Kwansei Gakuin University, Sanda Hyogo 669-1337, Japan}
\author{Hideki Hashimoto}
 \affiliation{School of Science and Technology, Kwansei Gakuin University, Sanda Hyogo 669-1337, Japan}






\begin{abstract}
We report on the sensing stability of quantum nanosensors in aqueous buffer solutions for the two detection schemes of quantum decoherence spectroscopy and nanoscale thermometry. The electron spin properties of single nitrogen-vacancy (NV) centers in 25-nm-sized nanodiamonds have been characterized by observing individual nanodiamonds during a continuous pH change from 4 to 11.
We have determined the stability of the NV quantum sensors during the pH change as the fluctuations of $\pm$12\% and $\pm$0.2 MHz for the spin coherence time ($T_2$) and the resonance frequency ($\omega _0$) of their mean values, which are comparable to the instrument error of the measurement system.
We discuss the importance of characterizing the sensing stability during the pH change and how the present observation affects the measurement scheme of nanodiamond-based NV quantum sensing.

\end{abstract}

\maketitle

\section{Introduction}

Diamond nitrogen-vacancy (NV) centers have attracted much attention as nanoscale quantum sensors~\cite{balasubramanian2008nanoscale,Doherty20131,maze2008nanoscale,plakhotnik2015super,hemelaar2017nanodiamonds,ong2017shape,fujiwara2016manipulation,fujiwara2015ultrathin,fujiwara2017fiber}.
NV centers possess unpaired electron spins in diamond lattice structures that can be optically detected (optically detected magnetic resonance: ODMR) with ultra-high sensitivity down to the single electron spin level
~\cite{jelezko2006single,Hanson2006,childress2006coherent}. 
These electron spin properties, such as the resonance frequency and the spin relaxation time, are dependent on 
multiple physical quantities (magnetic field, electric field, and temperature), so that one can measure their local values around the NV quantum sensors
~\cite{balasubramanian2008nanoscale,Doherty20131,maze2008nanoscale,plakhotnik2015super,hemelaar2017nanodiamonds,ong2017shape,fujiwara2016manipulation}.

This attractive multifunctionality of the NV quantum sensors, however, complicates data analysis under most physiological conditions.
Among the NV-quantum sensing schemes, 
quantum decoherence spectroscopy~\cite{hall2010monitoring,cole2009scanning,mcguinness2011quantum,luan2015decoherence} and nanoscale thermometry~\cite{kucsko2013nanometre,neumann2013high,tzeng2015time,simpson2017non} are promising for biological applications.
In decoherence spectroscopy, the change in the $T_2$ coherence time is detected, while, in thermometry, the frequency shift of the electron spin resonance is observed.
These spin properties (either $T_2$ or resonance frequency $\omega_0$) can be simultaneously affected by various parameters, for example, local heat generation in cells~\cite{kucsko2013nanometre,neumann2013high,donner2012mapping,okabe2012intracellular,liu2015intracellular,tsuji2017difference} or the local concentrations of ions~\cite{hall2010monitoring} and pH~\cite{tsou2015local,morimoto2016high}.
Thus, before proceeding to real biological applications, one needs experiments under controlled conditions to exclude factors, other than the sensing target, that may change the spin properties. 

Nanodiamonds are biocompatible and are excellent NV carriers that can be delivered into complex biological structures including cells, organelles, and tissues~\cite{simpson2017non,beranova2014sensitivity,turcheniuk2017biomedical,mohan2010vivo}. 
This portability provides distinct advantages over bulk-diamond-NV centers in biological sensing.
However, the spin properties of nanodiamond NV centers are more sensitive to the surface chemistry than the bulk-diamond NV centers because of the small distance between the NV centers and the surface, which degrades the spin properties~\cite{romach2015spectroscopy,ishikawa2012optical,rondin2010surface,song2014statistical}.
Moreover, the surface of nanodiamonds is neither uniform nor well defined in contrast to the flat surface of bulk diamond~\cite{kaur2013nanodiamonds,nagl2015improving,girard2011surface,paci2013understanding}. 
The spin properties of nanodiamond NV centers are, thus, thought to be more susceptible in physiological conditions.

One of the most influential parameters in biochemical experiments is the pH.
As chemical sensors, nanodiamonds are required to be used in a wide range of pH~\cite{rendler2017optical,schafer2001redox,slegerova2015designing}.
For intracellular applications, nanodiamonds will experience various pH depending on the locations; for example of cellular uptake, 
endosomes show pH of 6.0-7.0 in the early stage of endocytosis and later around 4.0~\cite{sorkin2002signal}.
The pH affects the surface potential of the nanodiamonds and has been demonstrated to convert the charge state of the NV centers between NV$^0$ and NV$^-$~\cite{karaveli2016modulation,schreyvogel2015active,ji2016charge}. 
Very recently there have been several reports that this charge-state instability indeed affects the ODMR measurements, such as compromising  the measured spin relaxation time~\cite{yamano2017charge,stacey2018evidence,bluvstein2019identifying}.
Thus properly characterizing the spin coherence of the ND quantum sensors in various pH conditions is required.

Here, we report on the spin-coherence stability of single nanodiamond-NV quantum sensors during a continuous pH change between 4 and 11, the range that is of particular importance for biochemical experiments.
During the pH change, the NV spin coherence time ($T_2$) and the spin resonance frequency ($\omega_0$) did not show any particular dependence on the pH but fluctuations corresponding to that observed in the steady-pH buffer solutions.
We have determined the sensor stability of the NV quantum systems, which provides the fluctuations of $\pm$12\% and $\pm$0.2 MHz for $T_2$ and $\omega _0$ of their mean values.
The observed fluctuations are discussed in relation to the NV-quantum sensing schemes such as decoherence spectroscopy and thermometry.

\begin{figure}[t!]
\centering
\includegraphics[width=70mm]{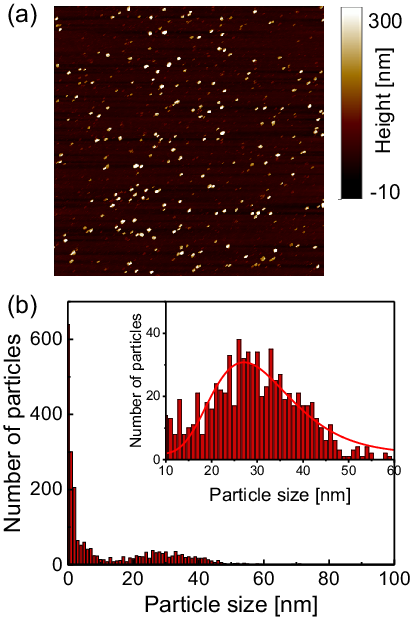}
\caption{(a) Atomic force microscopy topography image of $8\times 8$ \si{\um} region and (b) the corresponding particle size distribution. 
A large number of particles smaller than 10 nm in the region are considered to be debris (not diamond) included in the centrifugation process.
}
\label{fig0}
\end{figure}

\begin{figure*}[t!]
\centering
\includegraphics[width=170mm]{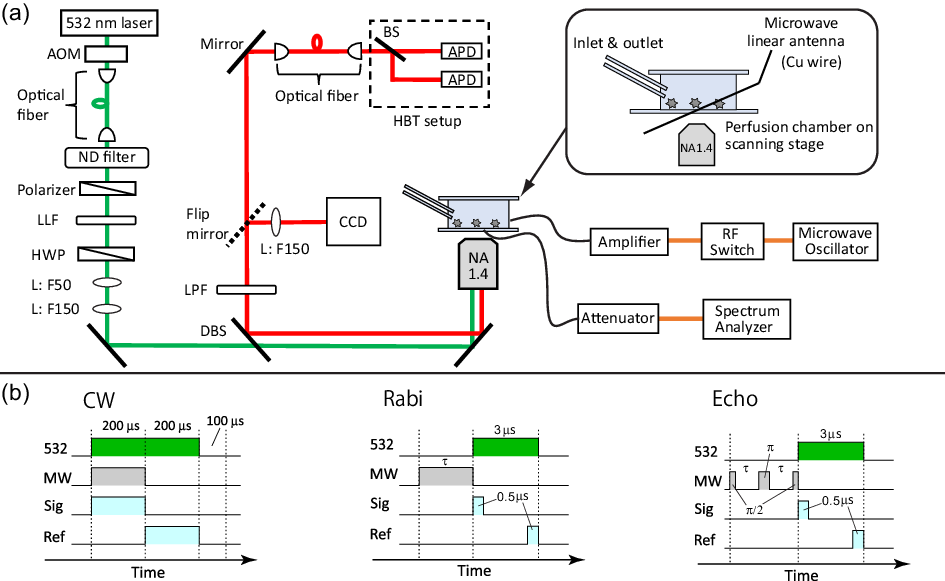}
\caption{(a) Schematic drawings of the experimental setup for the perfusion chamber,  optical layout and microwave circuit.
AOM: acousto-optic modulator. ND: neutral density: LLF: laser line filter. HWP: half-wave plate. L: lens. DBS: dichroic beam splitter. LPF: long pass filter. CCD: charge-coupled device camera. BS: beam splitter. APD: avalanche photodiode.
(b) The pulse sequences used for the electron spin measurements for CW, Rabi, and echo. 532: green laser. MW: microwave. Sig: signal counter. Ref: reference counter. }
\label{fig1}
\end{figure*}

\section{Experiments}
\subsection{Sample preparation}
A commercially available nanodiamond suspension (Microdiamant, MSY 0-0.05, median particle size: 25 nm) was purified by centrifugation and dispersed in distilled water.
A small droplet of the suspension was spin-coated on a cleaned coverslip to disperse and immobilize the nanodiamond particles on the coverslip surface. 
A 25-\si{\um}-thin copper wire was placed on the coverslip and both of the ends were soldered with electric connectors.
An acrylic chamber with a height of a few millimeters with inlet and outlet tubes was then glued on the coverslip using a UV-curing resin. 
It was sealed with a glass plate.

The topographies of the spin-coated samples were obtained using an atomic force microscope (AFM, Bruker, Edge). The AFM image is shown in Fig.~\ref{fig0}a. 
The peak heights of the distributed nanodiamonds were measured to obtain the particle size distribution, which indicated a mean particle size of 30 nm based on log-normal fitting (Fig.~\ref{fig0}b). 
Dynamic light scattering measurements (Malvern, Zetasizer Nano ZS) were also used to obtain the particle size distribution, which indicates a size of 58 nm (log-normal fitting), which is slightly larger than the result obtained from the AFM measurements (Fig.~\ref{fig0}c).

\subsection{Optical measurements}
The perfusion chamber was mounted on a three-axis piezo stage and observed by a home-built confocal fluorescence microscope (Fig.~\ref{fig1}a).
A continuous-wave 532-nm laser was used for the excitation with a typical excitation intensity of ca. 90 kW$\cdot {\rm cm ^{-2}}$ for the image scanning and second-order photon correlation measurements, 
which is near the fluorescence saturation laser intensity (Supplementary Fig.~S1$^{\dag}$).
An oil-immersion microscope objective with numerical aperture of 1.4 
was used for both the excitation and the fluorescence collection. 
The NV fluorescence was filtered by a dichroic beam splitter (Semrock, FF560-FDi01) 
and a long pass filter (Semrock, BLP01-561R) to remove the residual green laser scattering. 
The microscope was then coupled to an optical fiber that acted as a pinhole (Thorlabs, 1550HP, core diameter ca. 10 \si{\um}).
The fiber-coupled fluorescence was finally guided into a Hanbury--Brown--Twiss (HBT) setup that consisted of two avalanche photodiodes (Perkin Elmer SPCM AQRH-14) and a 50:50 beam splitter. 
For the spectral measurements, the microscope was connected to a fiber-coupled spectrometer equipped with a liquid-nitrogen cooled charge-coupled device (CCD) camera (Princeton, LNCCD). 
By scanning the sample with the piezo stage, we were able to obtain fluorescence scanning images of the nanodiamonds.
A time-correlated single-photon counting module (PicoQuant, TimeHarp-260) was used to obtain second-order photon correlation histograms to identify single NV centers by measuring the antibunching.

The perfusion chamber was first filled with water and then exchanged with the buffer solutions. 
We used two kinds of mixed buffer solutions to measure the pH range between 4--7 and 7--11. 
A citric acid (0.1 M)--Na$_2$HPO$_4$ (0.2 M) mixed buffer solution was used 
to control the pH range to 4--7.  
For the pH range of 7--11, a Na$_2$CO$_3$ (0.1 M)--HCl (0.5 M) mixed solution was used. 
The pH of the solutions was varied stepwise by $\Delta {\rm pH} \sim 1$ by changing the mixing ratio of the two constituents.
During the optical excitation, these solutions were pumped through the perfusion chamber continuously at a rate of 80 $\si{\uL} \cdot {\rm min}^{-1}$ to prevent photothermal aggregation of the  nanoparticles~\cite{nishimura2014control}
(these nanoparticles
may be nanodiamonds detached from other places or ionic salt
nanocrystals created by mixing the buffer solutions).

\subsection{Electron spin resonance (ESR) measurements.}
Microwaves were generated from a microwave source (Rohde \& Schwarz, SMB100A) and amplified by 45 dB (Mini-circuit, ZHL-16W-43+). The microwaves were fed to the microwave linear antenna in the perfusion chamber (Fig.~\ref{fig1}a).
The typical microwave excitation power for the continuous-wave ODMR spectral measurement was 35 dBm (3.2 W).
The avalanche photodiode (APD) detection was gated for microwave irradiation ON and OFF 
by using a radiofrequency (RF) switch (Mini-circuit, ZYSWA-2-50DR-S) and a bit pattern generator (Spincore, PBESR-PRO-300)~\cite{geiselmann2013three}. 
The gate width was 200 \si{\us}, common to both gates, followed by a laser shut-off time of 100 $\si{\us}$, 
giving $I_{PL}^{ON}$ and $I_{PL}^{OFF}$ (see Fig.~\ref{fig1}b) with a repetition rate of 2 kHz.
Note that an external magnetic field was not applied in this study. 
We selected NV centers that showed naturally single or doubly split (and well separated in frequency) ODMR peaks to excite only the single resonance peak of either of the transitions between $\ket{0} \to \ket{\pm 1}$ in the following pulsed ODMR measurements.~\cite{liu2013fiber,tisler2009fluorescence}.
The Rabi and spin echo measurements were performed on either of $\ket{0} \to \ket{\pm 1}$ transitions identified from the cw  ODMR spectra.
The Rabi signal determines the pulse durations of $\pi /2$ and $\pi$ pulses for the subsequent spin echo measurements (see Fig.~\ref{fig1}d). 
The spin echo measurements determine the spin coherence time ($T_2$). 
Note that we measured both $\pi /2$--$\pi$--$\pi /2$ and $\pi /2$--$\pi$--$3\pi /2$ sequences and subtracted these signals from each other to cancel the common mode noise in the spin echo measurements~\cite{bolshedvorskii2017single}.

\begin{figure*}[t!]
\centering
\includegraphics[width=160mm]{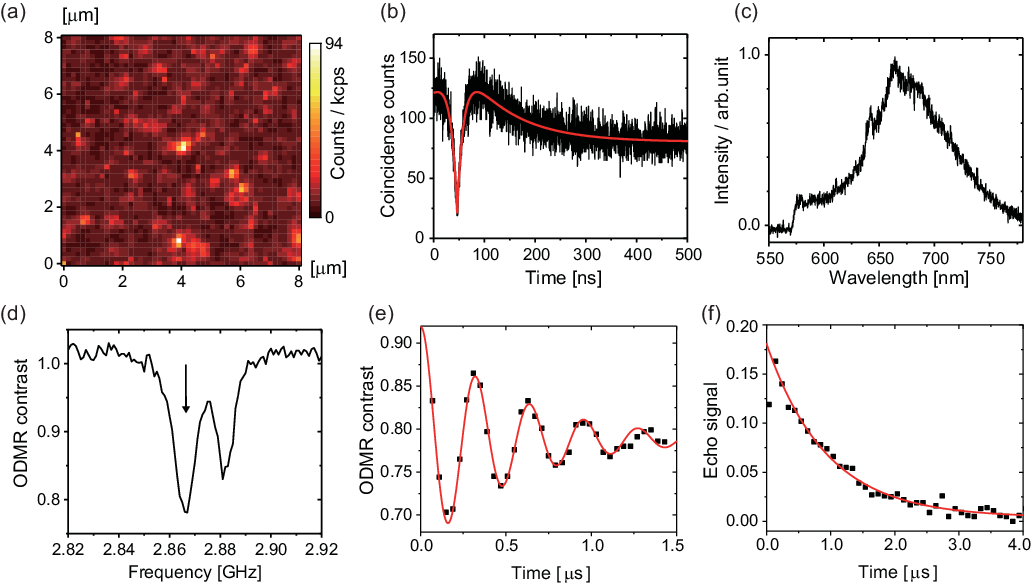}
\caption{(a) A confocal fluorescence scanning image of the nanodiamonds in the citric-acid--Na$_2$HPO$_4$ buffer solution at pH 7.0. (b) A second-order fluorescence photon correlation histogram of the central fluorescence spot in Fig.~\ref{fig2}a. The accumulation time was 100 s. (c) The fluorescence spectrum of the emission. (d) The ODMR spectrum of the NV center. (e) Its Rabi profile and (f) spin-echo profile. 
The Rabi and spin echo measurements were performed on the $\ket{0} \to \ket{-1}$ transition (indicated by the arrow) identified from the cw-ODMR spectrum in this measurement.}
\label{fig2}
\end{figure*}

\section{Results and Discussion}
\begin{figure}[t!]
\centering
\includegraphics[width=75mm]{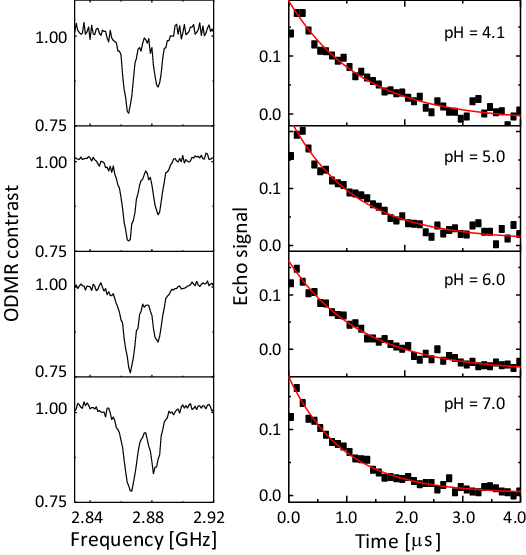}
\caption{The CW-ODMR spectra (left panel) and the corresponding spin-echo profiles (right panel) of the single NV center during a pH change from 4.1 to 7.0.}
\label{fig3}
\end{figure}

Figure~\ref{fig2}a shows a confocal fluorescence scanning image of the nanodiamonds in the citric-acid--Na$_2$HPO$_4$ buffer solution of  pH 7.0; there is a fluorescent spot at the center.
Figures~\ref{fig2}b,c show the second-order photon correlation histogram and the spectrum of the emitted fluorescence, respectively. 
By curve fitting to the data based on the equation in Ref.~\citenum{berthel2015photophysics}, we determined the excited-state lifetime to be 13.0 ns and the lifetime of the nearby metastable state to be 95.8 ns. 
The fluorescence spectrum consists of a zero-phonon line at around 637 nm and a broad phonon sideband ranging from 650 to 750 nm.
These observed fluorescence properties allowed us to clearly identify nanodiamonds that incorporate single NV centers for the subsequent measurements.

After we had identified single NV centers, we measured the ODMR signals.
Figure~\ref{fig2}d shows the ODMR spectrum of a single NV center.
The peak splits without an external magnetic field because of the lattice strain in the nanodiamond, which is well known for NV centers in nanodiamonds~\cite{liu2013fiber,tisler2009fluorescence}. 
We set the microwave frequency to either of the two peaks (here, the left peak is chosen: 2.8660 GHz). 
Figures~\ref{fig2}e,f are temporal profiles of the Rabi and spin echo sequences.
In the Rabi sequence, the microwave pulse duration was varied and the resultant fluorescence change was recorded. By fitting the damped sinusoidal function to the data, we determined the time duration of the $\pi$ pulse. 
With this $\pi$ pulse (and its half pulse $\pi/2$), the spin echo can be measured. 
The spin echo signal shows exponential decay with a spin coherence time ($T_2$) of 1862 ns. 
We performed this set of the spin measurements each time that the pH was changed. 
Note that the first dot (20--40 ns) of the spin echo signal is omitted from the curve fitting because the pulse duration is not as short as designed because of the timing jitter of the RF switch (ca. 10 ns). 
Note also that we used a single exponential fitting to all the following data because of the short $T_2$ time. 
It is well known that spin echo profiles show $\exp{[-(2t/T_2)^\alpha]}$ with $\alpha = 1-3$ when $T_2$ is long, such as 10 \si{\us}~\cite{knowles2014observing,tisler2009fluorescence,mcguinness2013ambient}.
However, the echo profile can be approximated as a single exponential when $T_2$ time is short. 
We, therefore, use single exponential fitting by taking $\alpha$ = 1 in the present study.

\begin{figure*}[t!]
\centering
\includegraphics[scale=0.7]{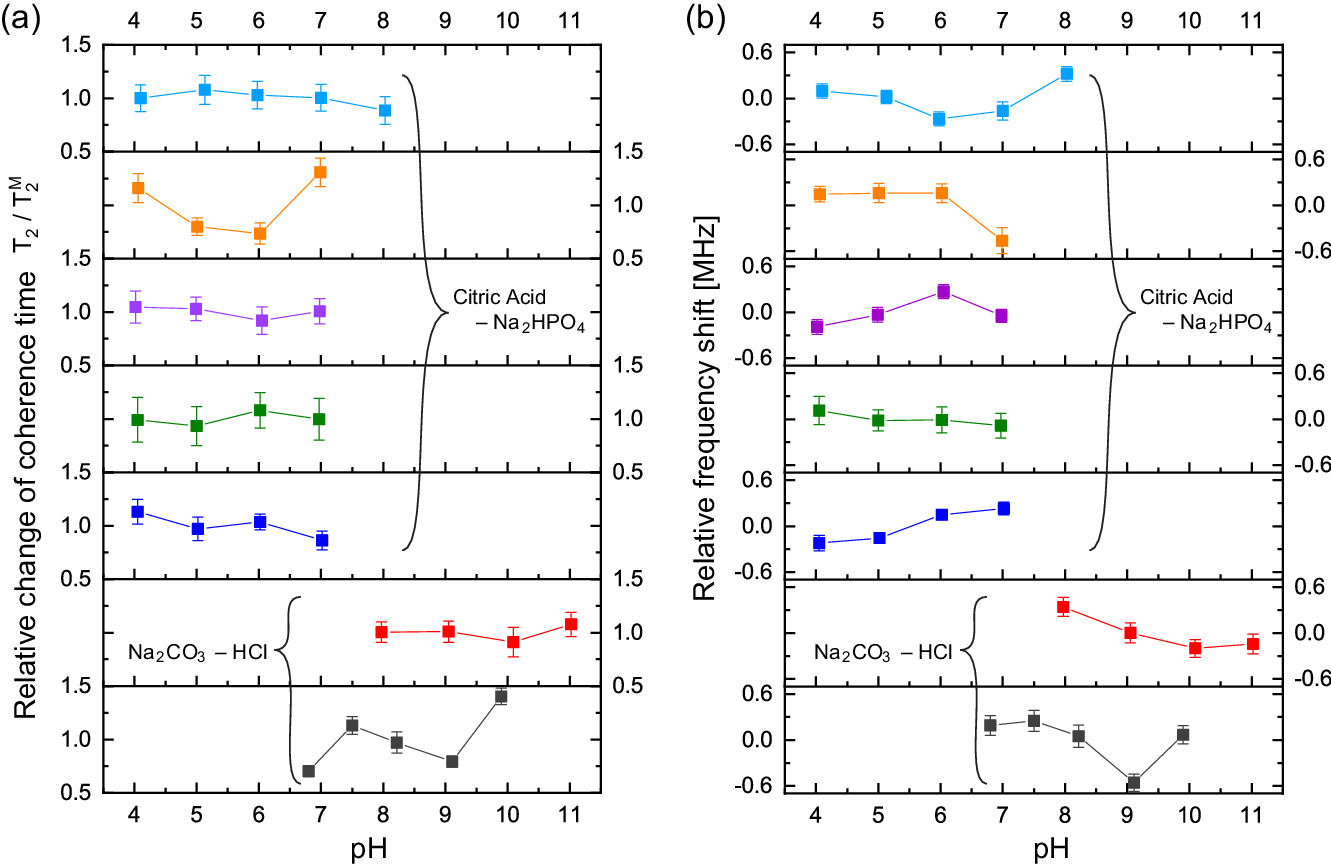}
\caption{(a) The fluctuation of each $T_2$ measurement value relative to its mean ($T_2^M$) as a function of pH. In total, seven nanodiamonds were investigated. All the nanodiamonds contain single NV centers. The upper-left five nanodiamonds were measured  in the phosphate-based buffer solution and the lower right two were in the carbonate-based buffer solution. The solid lines are for ease of visualization. (b) The corresponding $\omega _0$  fluctuation of the respective NV center to its mean over the pH range. The error bars are the fitting errors.}
\label{fig4}
\end{figure*}

Figure~\ref{fig3} shows the ODMR spectra and the corresponding echo profiles of the single NV center at pH 4 to 7. 
In the spin echo measurements, we excited the left peak.
The ODMR spectra are almost the same during the pH change and 
the echo profiles basically show single exponential decay. 
The $T_2$ times are 2491, 2202, 2495, and 1910 ns at pH 4.1, 5.0, 6.0, and 7.0, respectively. 
There was a fluctuation in the $T_2$ coherence time, but there is no clear dependence on the pH.
We, therefore, measured the echo profiles at different pH values for a number of single NV centers in the nanodiamonds to assess the pH dependence by statistical means. 
We measured the $T_2$ coherence time at each pH and normalized the value to the mean value ($T_2^M$) to show the relative deviation of the measurement to the mean ($T_2 / T_2^M$).
The results are graphically summarized in Fig.~\ref{fig4}a. 
Analogously, the resonance frequency was measured to determine the fluctuating peak shift from its mean value (Fig.~\ref{fig4}b).
Both the $T_2$ time and the resonance frequency  show a certain range of fluctuations with typical errors of 5--15 \% for $T_2$ and 0.06--0.18 MHz for $\omega _0$, but there is no clear dependence on the pH. 
The optical excitation power used in these measurements were 18--50 kW$\cdot {\rm cm ^{-2}}$.

\begin{figure}[th!]
\centering
\includegraphics[width=75mm]{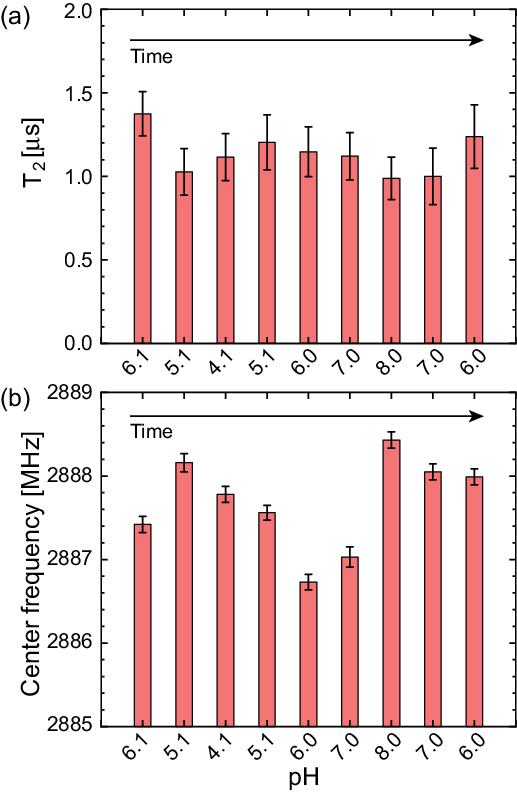}
\caption{(a) $T_2$ and (b) $\omega_0$ of a single NV center during the repeated pH change of 6.1 $\to$ 4.1 $\to$ 8.0 $\to$ 6.0, measured with a lower optical excitation intensity of 5.4 kW$\cdot {\rm cm ^{-2}}$. The error bars are the fitting errors.}
\label{figx1}
\end{figure}

In addition to these statistical measurements, we measured the NV sensing stability and repeatability in the same pH range with lower optical excitation intensity of 5.4 kW$\cdot {\rm cm ^{-2}}$ to avoid associated effects of the high optical excitation intensity, such as optical decoherence (laser power broadened linewidth) and photoionization of the  NV charge state, since these effects might affect the NV sensing stability determined in the above experiments. 
Figures~\ref{figx1} a,b show $T_2$ and $\omega_0$ of a single NV center during the repeated pH change from 6.1 $\to$ 5.1 $\to$ 4.1 $\to$ 5.1 $\to$ 6.0 $\to$ 7.0 $\to$ 8.0 $\to$ 7.0 $\to$ 6.0 with a step size of $\Delta$pH = 1. 
The data again show no significant dependence of the NV sensing ability on the pH change but show a certain range of fluctuations. 
While $T_2$ shows almost constant value within the error range, $\omega_0$ exhibits a relatively large shift compared with its error on the order of hundreds of kHz up to 1 MHz.

To clarify the origin of the fluctuation, 
we quantified the experimental error of the present quantum spin measurement scheme.
We measured the fluctuations of the $T_2$ time and resonance frequency over 19 h 
for single NV centers in the same buffer solution at constant pH of 6.1 with the similar optical excitation intensity of 5.4 kW$\cdot {\rm cm ^{-2}}$.
It was observed that $\omega_0$ shows sudden shifts on the order of hundreds of kHz as can be seen in Fig.~\ref{figx1}. 
The standard deviation of this 19-h measurement is $\pm 6.2$ \% of the mean $T_2$ time and $\pm$ 0.24 MHz of the mean $\omega_0$ (see Table~\ref{tbls1}).
Nevertheless, most of the data points shown in Fig.~\ref{fig4} are located within the range of this measurement fluctuation, 
indicating that the pH does not affect the $T_2$ and $\omega_0$ within the error range of the present spin measurement parameters (number of accumulation, step size, etc...).
It is therefore concluded that the effect of pH to $T_2$ and $\omega_0$ is smaller than the instrument measurement error of the present experimental system.
Note that the stability of the NV spin properties measured in the buffer solutions is almost the same (or even better) as that in an air environment (see Table~\ref{tbls2}). 

\begin{table}[t!]
  \caption{
    The spin properties of the single NV center in the phosphate-based buffer solution at pH 6.1 over 19 h.
  }
  \label{tbls1}
  \centering
  \begin{tabular}{ccccc}
	Time [h]  & $\omega_0^{\rm R}$ [GHz] & $\Delta \omega_0^{\rm R}$ [MHz]& $T_2$ [ns] & $\Delta T_2$ [ns] \\
    \hline
    0   			&	2.88757	& 0.10	& 1381	& 188 \\
    2.03 			&	2.88766	& 0.09	& 1251	& 143 \\
    5.8   			&	2.88898	& 0.12	& 1060	& 120 \\
    13.4  			&	2.88856	& 0.09	& 1343	& 142 \\
    17.4 			&	2.88837	& 0.10	& 937	& 158 \\
    19.1			&	2.88742	& 0.10	& 1374	& 133 \\
    \hline 
    Mean			&	2.88809	&		&1224	&	 \\
    Std. Err.		&	0.24 [MHz]	&	&76 [ns]	&  \\
      \hline
  \end{tabular}
\end{table}

\begin{table}[t!]
  \caption{
  The spin properties of the single NV center in air (on coverslip) over 24 h.}
  \label{tbls2}
  \centering
  \begin{tabular}{ccccc}
	Time [h]  & $\omega_0$ [GHz] & $\Delta \omega_0$ [MHz]& $T_2$ [ns] & $\Delta T_2$ [ns] \\
    \hline
    0   			&2.88758	&0.06		& 620	&81 \\
    4.42 			&2.88763	&0.11		& 531	&47 \\
    17.2   			&2.88830	&0.12		& 978	&102 \\
    20.3  			&2.88857	&0.14		& 916	&92 \\
    23.25 			&2.88916	&0.09		& 504 	&33 \\
    24.95			&2.88934	&0.01		& 420	&42 \\
    \hline 
    Mean			&2.88843	&			&662	& \\
    Std. Err.		&0.30 [MHz]	&			&94 (14\%)	&  \\
    \hline
  \end{tabular}
\end{table}

It should be noted that the fluctuations of $T_2$ and $\omega_0$ mainly come from the fluctuations of the fluorescence photon counts and the ODMR spectral shape. 
The pH change and aqueous environment affect the NV charge stability and hence fluorescence emission properties. 
Since the NV spin measurement relies on the fluorescence detection, these fluorescence fluctuation significantly affects the sensitivity and precision of the NV quantum sensors, i.e. making noise. 
While we have not observed any particular pH dependence of the NV quantum sensors, this noise can be regarded as the effect of pH change or buffer solutions to the NV quantum sensing.

Indeed, during the long-term optical excitation, some NV centers were destabilized. 
These NV centers showed stable fluorescence initially but were later destabilized, resulting in fluorescent bursts or 
blinking, as shown in Fig.S2. 
Such an instability is caused by the optical excitation in aqueous solutions and has not been observed in air, 
which causes the measurement noise in the end. 
We believe that this destabilization is related to the surface adsorption of water or solvent molecules, 
causing the photoionization of NV charge states. 
This point is important when performing the spin measurements in aqueous environment, and the continuous irradiation by the laser no longer guarantees the stability of the NV-fluorescent probes, particularly at the single NV level.
The effect of the photoionization needs to be reduced by decreasing the laser power in case of the long-term tracking of nanodaimonds, though robustness of the NV centers is different particle by particle (see Fig.~S3$^\dagger$ for example).

Increasing the surface oxidation coverage can also improve the photostability.
The photoionization is related to the charge instability of NV centers, which switches between NV$^0$ and NV$^-$. 
Such charge state conversion is strongly related to the surface termination of nanodiamonds, as has been demonstrated in relation to the charge conversion between these two states by applying an electric potential and surface termination~\cite{karaveli2016modulation,schreyvogel2015active,ji2016charge,yamano2017charge,hauf2011chemical}. 
The surface-oxidized diamonds do not show charge conversion under the electric field because the band is lower than the potential of the electrolyte. 
The surfaces of our present nanodiamonds could be oxidized further by acid cleaning or high-temperature treatment to prevent destabilization in aqueous solutions~\cite{stehlik2015size,wolcott2014surface}.
Note that measurement of the zeta potential of the nanodiamonds in the present experimental conditions is 
not possible because high ionic strength of the present buffer solutions causes the sedimentation of nanodiamonds~\cite{gibson2009colloidal,zhao2011chromatographic,zhang2011overview,barreto2015behavior}.
Zeta potential measurements in diluted suspensions are possible and performed as described in Fig. S4$^\dagger$.

We also note that the present experiment does not provide information on the relationship between the surface pH and NV spin properties. 
The surface inhomogeneity by which the nanodiamond-NV centers might have some pH dependence could result in significant differences between the surface pH and the bulk pH.
The surface pH is an important parameter for nanoparticle science~\cite{bellucci2008nanoparticles} in biological applications, and it may be necessary to study the effect of the surface pH in the future.

The present results concerning the fluctuations of $T_2$ and $\omega_0$ in the quantum spin measurement schemes 
provide an important direction for the real implementation of nanoscale thermometry~\cite{kucsko2013nanometre,neumann2013high,tzeng2015time,simpson2017non} and quantum decoherence spectroscopy~\cite{hall2010monitoring,cole2009scanning,luan2015decoherence}.
For example, NV-nanodiamond thermometry is promising for biological analysis and is expected to allow the observation of cells in activated conditions, such as neuronal excitation~\cite{barry2016optical,simpson2017non,tanimoto2016detection} and mitochondrial activity~\cite{okabe2012intracellular}.
The resonance frequency of ODMR has a temperature dependence of $-74$ kHz$\cdot {\rm K}^{-1}$ and, through this change, one can measure the local temperature around the nanodiamonds. 
A realistic temperature range monitored during mammalian cellular activities is 34--42 $^\circ$C with a standard temperature of 38 $^\circ$C~\cite{donner2012mapping,okabe2012intracellular,liu2015intracellular,tsuji2017difference}.
This means that only a temperature change of $\pm$4 $^\circ$C ($\pm$0.3 MHz), at largest, is expected.

Since our measurement method based on the curve fitting to the whole spectral profile is useful as a first choice
Given our experimental error of $\omega_0$ to be $\pm$0.2 MHz over the 16-h measurement, 
more stable and high precision schemes to determine $\omega_0$ with a fast acquisition time should be implemented.
While the present method based on curve-fitting to the whole spectral profile is useful as a first choice in the biological applications, 
it takes a long measurement time to reach the sufficient precision. 
With the data acquisition time of 12 min, it can reach an error of $\pm$ 35 kHz for $\omega_0$ (Fig.S4), 
though it is not realistic to spend more than 12 min to determine the parameters. 
Ratiometric analysis of the ODMR spectral shape may be a good choice to simultaneously obtain the precision and the fast acquisition speed. 
Another technique that may overcome this measurement error is a use of statistical analysis of number of NV nanodiamonds to obtain high precision~\cite{simpson2017non}. 
The same discussion also applies to the decoherence spectroscopy where the $T_2$ relaxation time is fitted to the whole profile of the exponential decay of the spin coherence time. 
In case of the decoherence spectroscopy, single point analysis has been proposed and demonstrated for $T_1$ measurements~\cite{simpson2017electron}. 
Implementing these ratiometric analyses or statistical analysis seem necessary for the real biological applications.

\section{Conclusion}
In conclusion, we have reported the effects of aqueous buffer solutions on the electron spin properties of single NV centers in 25-nm-sized nanodiamonds by tracking individual nanodiamonds during continuous pH changes in the range of pH 4 to 11.
During the pH changes, the $T_2$ time and the spin resonance frequency did not show any particular pH dependence but did show fluctuations that correspond to the experimental errors observed in the steady-pH buffer solutions 
over 19 h ($\pm$ 23\% and $\pm$ 0.2 MHz for $T_2$ and $\omega _0$ of their mean values). 
The observed fluctuations are significant when performing  nanoscale thermometry and quantum decoherence spectroscopy in various biological contexts and more stable and faster measurement schemes should be necessary such as ratio-metric analyses.

Measuring the stability of the spin properties of single nanodiamond-NV centers during the pH change is important to the future development of nanodiamond-based NV quantum sensing, because the measured spin properties can be compromised by the charge-state instability of NV centers as recently reported~\cite{bluvstein2019identifying}.
Furthermore, bare nanodiamonds cannot be dispersed in high-ionic-strength buffer solutions~\cite{gibson2009colloidal,zhao2011chromatographic} that include most biochemical buffers, such as phosphate-buffered-saline cell culture media. 
Consequently, it is important to functionalize the nanodiamond surface to prevent aggregation or binding with other biological molecules~\cite{genjo2016nanodiamond,sotoma2016selective}. 
Surface functionalization is also used for nanodiamond-based pH sensors~\cite{rendler2017optical}.  
However, after the surface functionalization, nanodiamonds must maintain the original spin properties at various pH values.
The present results and measurement scheme can provide a way to evaluate the stability of such quantum sensors in the stage of material development before the use, thus allowing the exploration of the applications of NV quantum sensors for biological and biochemical applications.

\section*{Conflicts of interest}
The authors declare no competing financial interests. 

\section*{Acknowledgements}
We thank Prof. Noboru Ohtani for the AFM measurements and 
Ms. Kaori Kobayashi and Mr. Yuta Ueda for the support in zeta-potential measurements.
MF acknowledges financial support from JSPS-KAKENHI (Nos. 26706007, 26610077, and 17H02741), MEXT-LEADER program, and Osaka City University (OCU-Strategic Research Grant 2017 for young researchers).
MF, HY thank JSPS-KAKENHI (No. 16K13646).
SS acknowledges financial support from JSPS-KAKENHI (No. 26220602).
HH thanks JSPS KAKENHI, Grant-in-Aids for Basic Research (B) (No. 16H04181) and Scientific Research on Innovative Areas "Innovations for Light-Energy Conversion (I4LEC)" (Nos. 17H06433, 17H0637) for financial support.


\begin{thebibliography}{10}

\bibitem{balasubramanian2008nanoscale}
Gopalakrishnan Balasubramanian, IY~Chan, Roman Kolesov, Mohannad Al-Hmoud,
  Julia Tisler, Chang Shin, Changdong Kim, Aleksander Wojcik, Philip~R Hemmer,
  Anke Krueger, Hanke Tobias, Alfred Leitenstorfer, Rudolf Bratschitsch, Fedor
  Jelezko, and J\"{o}rg Wrachtrup.
\newblock Nanoscale imaging magnetometry with diamond spins under ambient
  conditions.
\newblock {\em Nature}, 455(7213):648--651, 2008.

\bibitem{barreto2015behavior}
{\^A}ngela Barreto, Luis~G Luis, Ana~V Gir{\~a}o, Tito Trindade, Amadeu~MVM
  Soares, and Miguel Oliveira.
\newblock Behavior of colloidal gold nanoparticles in different ionic strength
  media.
\newblock {\em Journal of Nanoparticle Research}, 17(12):493, 2015.

\bibitem{barry2016optical}
John~F Barry, Matthew~J Turner, Jennifer~M Schloss, David~R Glenn, Yuyu Song,
  Mikhail~D Lukin, Hongkun Park, and Ronald~L Walsworth.
\newblock Optical magnetic detection of single-neuron action potentials using
  quantum defects in diamond.
\newblock {\em Proceedings of the National Academy of Sciences},
  113:14133--14138, 2016.

\bibitem{bellucci2008nanoparticles}
Stefano Bellucci.
\newblock {\em Nanoparticles and Nanodevices in Biological Applications: The
  INFN Lectures}, volume~1.
\newblock Springer Science \& Business Media, 2008.

\bibitem{beranova2014sensitivity}
Jana Beranov{\'a}, Gabriela Seydlov{\'a}, Halyna Kozak, Old{\v{r}}ich Benada,
  Radovan Fi{\v{s}}er, Anna Artemenko, Ivo Konop{\'a}sek, and Alexander Kromka.
\newblock Sensitivity of bacteria to diamond nanoparticles of various size
  differs in gram-positive and gram-negative cells.
\newblock {\em FEMS Microbiology Letters}, 351(2):179--186, 2014.

\bibitem{berthel2015photophysics}
Martin Berthel, Oriane Mollet, G{\'e}raldine Dantelle, Thierry Gacoin, Serge
  Huant, and Aur{\'e}lien Drezet.
\newblock Photophysics of single nitrogen-vacancy centers in diamond
  nanocrystals.
\newblock {\em Physical Review B}, 91(3):035308, 2015.

\bibitem{bluvstein2019identifying}
Dolev Bluvstein, Zhiran Zhang, and Ania C~Bleszynski Jayich.
\newblock Identifying and mitigating charge instabilities in shallow diamond
  nitrogen-vacancy centers.
\newblock {\em Physical Review Letters}, 122(7):076101, 2019.

\bibitem{bolshedvorskii2017single}
Stepan~V Bolshedvorskii, Vadim~V Vorobyov, Vladimir~V Soshenko, Vladimir~A
  Shershulin, Javid Javadzade, Anton~I Zeleneev, Sofya~A Komrakova, Vadim~N
  Sorokin, Peter~I Belobrov, Andrey~N Smolyaninov, and Alexey~V. Akimov.
\newblock Single bright nv centers in aggregates of detonation nanodiamonds.
\newblock {\em Optical Materials Express}, 7(11):4038--4049, 2017.

\bibitem{childress2006coherent}
L~Childress, MV~Gurudev Dutt, JM~Taylor, AS~Zibrov, F~Jelezko, J~Wrachtrup,
  PR~Hemmer, and MD~Lukin.
\newblock Coherent dynamics of coupled electron and nuclear spin qubits in
  diamond.
\newblock {\em Science}, 314(5797):281--285, 2006.

\bibitem{cole2009scanning}
Jared~H Cole and Lloyd~CL Hollenberg.
\newblock Scanning quantum decoherence microscopy.
\newblock {\em Nanotechnology}, 20(49):495401, 2009.

\bibitem{Doherty20131}
Marcus~W. Doherty, Neil~B. Manson, Paul Delaney, Fedor Jelezko, J\"{o}rg
  Wrachtrup, and Lloyd~C.L. Hollenberg.
\newblock The nitrogen-vacancy colour centre in diamond.
\newblock {\em Phys. Reports}, 528(1):1 -- 45, 2013.

\bibitem{donner2012mapping}
Jon~S Donner, Sebastian~A Thompson, Mark~P Kreuzer, Guillaume Baffou, and
  Romain Quidant.
\newblock Mapping intracellular temperature using green fluorescent protein.
\newblock {\em Nano Letters}, 12(4):2107--2111, 2012.

\bibitem{fujiwara2017fiber}
Masazumi Fujiwara, Oliver Neitzke, Tim Schr{\"o}der, Andreas~W Schell, Janik
  Wolters, Jiabao Zheng, Sara Mouradian, Mohamed Almoktar, Shigeki Takeuchi,
  Dirk Englund, et~al.
\newblock Fiber-coupled diamond micro-waveguides toward an efficient quantum
  interface for spin defect centers.
\newblock {\em ACS Omega}, 2(10):7194--7202, 2017.

\bibitem{fujiwara2016manipulation}
Masazumi Fujiwara, Kazuma Yoshida, Tetsuya Noda, Hideaki Takashima, Andreas~W
  Schell, Norikazu Mizuochi, and Shigeki Takeuchi.
\newblock Manipulation of single nanodiamonds to ultrathin fiber-taper
  nanofibers and control of nv-spin states toward fiber-integrated
  $\lambda$-systems.
\newblock {\em Nanotechnology}, 27(45):455202, 2016.

\bibitem{fujiwara2015ultrathin}
Masazumi Fujiwara, Hong-Quan Zhao, Tetsuya Noda, Kazuhiro Ikeda, Hitoshi
  Sumiya, and Shigeki Takeuchi.
\newblock Ultrathin fiber-taper coupling with nitrogen vacancy centers in
  nanodiamonds at cryogenic temperatures.
\newblock {\em Optics Letters}, 40(24):5702--5705, 2015.

\bibitem{geiselmann2013three}
Michael Geiselmann, Mathieu~L Juan, Jan Renger, Jana~M Say, Louise~J Brown,
  F~Javier~Garc{\'\i}a De~Abajo, Frank Koppens, and Romain Quidant.
\newblock Three-dimensional optical manipulation of a single electron spin.
\newblock {\em Nature Nanotechnology}, 8(3):175--179, 2013.

\bibitem{genjo2016nanodiamond}
Takuya Genjo, Shingo Sotoma, Ryotaro Tanabe, Ryuji Igarashi, and Masahiro
  Shirakawa.
\newblock A nanodiamond-peptide bioconjugate for fluorescence and odmr
  microscopy of a single actin filament.
\newblock {\em Analytical Sciences}, 32(11):1165--1170, 2016.

\bibitem{gibson2009colloidal}
N~Gibson, O~Shenderova, TJM Luo, S~Moseenkov, V~Bondar, A~Puzyr, K~Purtov,
  Z~Fitzgerald, and DW~Brenner.
\newblock Colloidal stability of modified nanodiamond particles.
\newblock {\em Diamond and Related materials}, 18(4):620--626, 2009.

\bibitem{girard2011surface}
HA~Girard, T~Petit, S~Perruchas, T~Gacoin, C~Gesset, JC~Arnault, and
  P~Bergonzo.
\newblock Surface properties of hydrogenated nanodiamonds: a chemical
  investigation.
\newblock {\em Physical Chemistry Chemical Physics}, 13(24):11517--11523, 2011.

\bibitem{hall2010monitoring}
Liam~T Hall, Charles~D Hill, Jared~H Cole, Brigitte St{\"a}dler, Frank Caruso,
  Paul Mulvaney, J{\"o}rg Wrachtrup, and Lloyd~CL Hollenberg.
\newblock Monitoring ion-channel function in real time through quantum
  decoherence.
\newblock {\em Proceedings of the National Academy of Sciences},
  107(44):18777--18782, 2010.

\bibitem{Hanson2006}
R.~Hanson, O.~Gywat, and D.~D. Awschalom.
\newblock {Room-temperature manipulation and decoherence of a single spin in
  diamond}.
\newblock {\em Physical Review B}, 74(16):1--4, 2006.

\bibitem{hauf2011chemical}
MV~Hauf, B~Grotz, B~Naydenov, M~Dankerl, S~Pezzagna, J~Meijer, F~Jelezko,
  J~Wrachtrup, M~Stutzmann, F~Reinhard, et~al.
\newblock Chemical control of the charge state of nitrogen-vacancy centers in
  diamond.
\newblock {\em Physical Review B}, 83(8):081304, 2011.

\bibitem{hemelaar2017nanodiamonds}
SR~Hemelaar, P~de~Boer, M~Chipaux, W~Zuidema, T~Hamoh, F~Perona Martinez,
  A~Nagl, JP~Hoogenboom, BNG Giepmans, and R~Schirhagl.
\newblock Nanodiamonds as multi-purpose labels for microscopy.
\newblock {\em Scientific Reports}, 7(720), 2017.

\bibitem{ishikawa2012optical}
Toyofumi Ishikawa, Kai-Mei~C Fu, Charles Santori, Victor~M Acosta, Raymond~G
  Beausoleil, Hideyuki Watanabe, Shinichi Shikata, and Kohei~M Itoh.
\newblock Optical and spin coherence properties of nitrogen-vacancy centers
  placed in a 100 nm thick isotopically purified diamond layer.
\newblock {\em Nano Letters}, 12(4):2083--2087, 2012.

\bibitem{jelezko2006single}
F~Jelezko and J~Wrachtrup.
\newblock Single defect centres in diamond: A review.
\newblock {\em Phys. Stat. Solidi (a)}, 203(13):3207--3225, 2006.

\bibitem{ji2016charge}
Peng Ji and MV~Gurudev Dutt.
\newblock Charge state dynamics of the nitrogen vacancy center in diamond under
  1064-nm laser excitation.
\newblock {\em Physical Review B}, 94(2):024101, 2016.

\bibitem{karaveli2016modulation}
Sinan Karaveli, Ophir Gaathon, Abraham Wolcott, Reyu Sakakibara, Or~A Shemesh,
  Darcy~S Peterka, Edward~S Boyden, Jonathan~S Owen, Rafael Yuste, and Dirk
  Englund.
\newblock Modulation of nitrogen vacancy charge state and fluorescence in
  nanodiamonds using electrochemical potential.
\newblock {\em Proceedings of the National Academy of Sciences},
  113(15):3938--3943, 2016.

\bibitem{kaur2013nanodiamonds}
Randeep Kaur and Ildiko Badea.
\newblock Nanodiamonds as novel nanomaterials for biomedical applications: drug
  delivery and imaging systems.
\newblock {\em International Journal of Nanomedicine}, 8:203, 2013.

\bibitem{knowles2014observing}
Helena~S Knowles, Dhiren~M Kara, and Mete Atat{\"u}re.
\newblock Observing bulk diamond spin coherence in high-purity nanodiamonds.
\newblock {\em Nature Materials}, 13(1):21--25, 2014.

\bibitem{kucsko2013nanometre}
G~Kucsko, PC~Maurer, NY~Yao, M~Kubo, HJ~Noh, PK~Lo, H~Park, and MD~Lukin.
\newblock Nanometre-scale thermometry in a living cell.
\newblock {\em Nature}, 500(7460):54--58, 2013.

\bibitem{liu2015intracellular}
Helin Liu, Yanyan Fan, Jianhai Wang, Zhongsen Song, Hao Shi, Rongcheng Han,
  Yinlin Sha, and Yuqiang Jiang.
\newblock Intracellular temperature sensing: An ultra-bright luminescent
  nanothermometer with non-sensitivity to ph and ionic strength.
\newblock {\em Scientific Reports}, 5:14879, 2015.

\bibitem{liu2013fiber}
Xiaodi Liu, Jinming Cui, Fangwen Sun, Xuerui Song, Fupan Feng, Junfeng Wang,
  Wei Zhu, Liren Lou, and Guanzhong Wang.
\newblock Fiber-integrated diamond-based magnetometer.
\newblock {\em Applied Physics Letters}, 103(14):143105--143105, 2013.

\bibitem{luan2015decoherence}
Lan Luan, Michael~S Grinolds, Sungkun Hong, Patrick Maletinsky, Ronald~L
  Walsworth, and Amir Yacoby.
\newblock Decoherence imaging of spin ensembles using a scanning
  single-electron spin in diamond.
\newblock {\em Scientific Reports}, 5, 2015.

\bibitem{maze2008nanoscale}
JR~Maze, PL~Stanwix, JS~Hodges, S~Hong, JM~Taylor, P~Cappellaro, L~Jiang,
  MV~Gurudev Dutt, E~Togan, AS~Zibrov, A.~Yacoby, R.~L. Walsworth, and M.~D.
  Lukin.
\newblock Nanoscale magnetic sensing with an individual electronic spin in
  diamond.
\newblock {\em Nature}, 455(7213):644--647, 2008.

\bibitem{mcguinness2011quantum}
Liam~P McGuinness, Yuling Yan, Alastair Stacey, David~A Simpson, Liam~T Hall,
  Dougal Maclaurin, Steven Prawer, P~Mulvaney, J~Wrachtrup, F~Caruso, R.~E.
  Scholten, and L.~C.~L. Hollenberg.
\newblock Quantum measurement and orientation tracking of fluorescent
  nanodiamonds inside living cells.
\newblock {\em Nature Nanotechnology}, 6(6):358, 2011.

\bibitem{mcguinness2013ambient}
LP~McGuinness, LT~Hall, A~Stacey, DA~Simpson, CD~Hill, JH~Cole, K~Ganesan,
  BC~Gibson, S~Prawer, P~Mulvaney, F.~Jelezko, J.~Wrachtrup, R.~E. Scholten,
  and L.~C.~L. Hollenberg.
\newblock Ambient nanoscale sensing with single spins using quantum
  decoherence.
\newblock {\em New Journal of Physics}, 15(7):073042, 2013.

\bibitem{mohan2010vivo}
Nitin Mohan, Chao-Sheng Chen, Hsiao-Han Hsieh, Yi-Chun Wu, and Huan-Cheng
  Chang.
\newblock In vivo imaging and toxicity assessments of fluorescent nanodiamonds
  in caenorhabditis elegans.
\newblock {\em Nano Letters}, 10(9):3692--3699, 2010.

\bibitem{morimoto2016high}
Yusuke~V Morimoto, Nobunori Kami-Ike, Tomoko Miyata, Akihiro Kawamoto, Takayuki
  Kato, Keiichi Namba, and Tohru Minamino.
\newblock High-resolution ph imaging of living bacterial cells to detect local
  ph differences.
\newblock {\em mBio}, 7(6):e01911--16, 2016.

\bibitem{nagl2015improving}
Andreas Nagl, Simon~Robert Hemelaar, and Romana Schirhagl.
\newblock Improving surface and defect center chemistry of fluorescent
  nanodiamonds for imaging purposes—a review.
\newblock {\em Analytical and Bioanalytical Chemistry}, 407(25):7521--7536,
  2015.

\bibitem{neumann2013high}
Philipp Neumann, Ingmar Jakobi, Florian Dolde, Christian Burk, Rolf Reuter,
  Gerald Waldherr, Jan Honert, Thomas Wolf, Andreas Brunner, Jeong~Hyun Shim,
  D~Suter, H~Sumiya, J~Isoya, and J~Wrachtrup.
\newblock High-precision nanoscale temperature sensing using single defects in
  diamond.
\newblock {\em Nano Letters}, 13(6):2738--2742, 2013.

\bibitem{nishimura2014control}
Yushi Nishimura, Keisuke Nishida, Yojiro Yamamoto, Syoji Ito, Shiho Tokonami,
  and Takuya Iida.
\newblock Control of submillimeter phase transition by collective photothermal
  effect.
\newblock {\em The Journal of Physical Chemistry C}, 118(32):18799--18804,
  2014.

\bibitem{okabe2012intracellular}
Kohki Okabe, Noriko Inada, Chie Gota, Yoshie Harada, Takashi Funatsu, and
  Seiichi Uchiyama.
\newblock Intracellular temperature mapping with a fluorescent polymeric
  thermometer and fluorescence lifetime imaging microscopy.
\newblock {\em Nature Communications}, 3:705, 2012.

\bibitem{ong2017shape}
SY~Ong, M~Chipaux, A~Nagl, and R~Schirhagl.
\newblock Shape and crystallographic orientation of nanodiamonds for quantum
  sensing.
\newblock {\em Physical Chemistry Chemical Physics}, 19(17):10748--10752, 2017.

\bibitem{paci2013understanding}
Jeffrey~T Paci, Han~B Man, Biswajit Saha, Dean Ho, and George~C Schatz.
\newblock Understanding the surfaces of nanodiamonds.
\newblock {\em The Journal of Physical Chemistry C}, 117(33):17256--17267,
  2013.

\bibitem{plakhotnik2015super}
Taras Plakhotnik, Haroon Aman, Shaohua Zhang, and Zhen Li.
\newblock Super-paramagnetic particles chemically bound to luminescent diamond:
  single nanocrystals probed with optically detected magnetic resonance.
\newblock {\em The Journal of Physical Chemistry C}, 119(34):20119--20124,
  2015.

\bibitem{rendler2017optical}
Torsten Rendler, Jitka Neburkova, Ondrej Zemek, Jan Kotek, Andrea Zappe, Zhiqin
  Chu, Petr Cigler, and J{\"o}rg Wrachtrup.
\newblock Optical imaging of localized chemical events using programmable
  diamond quantum nanosensors.
\newblock {\em Nature Communications}, 8:14701, 2017.

\bibitem{romach2015spectroscopy}
Y~Romach, C~M{\"u}ller, T~Unden, LJ~Rogers, T~Isoda, KM~Itoh, M~Markham,
  A~Stacey, J~Meijer, S~Pezzagna, et~al.
\newblock Spectroscopy of surface-induced noise using shallow spins in diamond.
\newblock {\em Physical Review Letters}, 114(1):017601, 2015.

\bibitem{rondin2010surface}
L~Rondin, G~Dantelle, A~Slablab, F~Grosshans, F~Treussart, P~Bergonzo,
  S~Perruchas, T~Gacoin, M~Chaigneau, H-C Chang, et~al.
\newblock Surface-induced charge state conversion of nitrogen-vacancy defects
  in nanodiamonds.
\newblock {\em Physical Review B}, 82(11):115449, 2010.

\bibitem{schafer2001redox}
Freya~Q Schafer and Garry~R Buettner.
\newblock Redox environment of the cell as viewed through the redox state of
  the glutathione disulfide/glutathione couple.
\newblock {\em Free Radical Biology and Medicine}, 30(11):1191--1212, 2001.

\bibitem{schreyvogel2015active}
C~Schreyvogel, V~Polyakov, R~Wunderlich, J~Meijer, and CE~Nebel.
\newblock Active charge state control of single nv centres in diamond by
  in-plane al-schottky junctions.
\newblock {\em Scientific Reports}, 5:12160, 2015.

\bibitem{simpson2017non}
David~A Simpson, Emma Morrisroe, Julia~M McCoey, Alain~H Lombard, Dulini~C
  Mendis, François Treussart, Liam~T Hall, Steven Petrou, and Lloyd~CL
  Hollenberg.
\newblock Non-neurotoxic nanodiamond probes for intraneuronal temperature
  mapping.
\newblock {\em ACS Nano}, 11:12077--12086, 2017.

\bibitem{simpson2017electron}
David~A Simpson, Robert~G Ryan, Liam~T Hall, Evgeniy Panchenko, Simon~C Drew,
  Steven Petrou, Paul~S Donnelly, Paul Mulvaney, and Lloyd~CL Hollenberg.
\newblock Electron paramagnetic resonance microscopy using spins in diamond
  under ambient conditions.
\newblock {\em Nature Communications}, 8(1):458, 2017.

\bibitem{slegerova2015designing}
Jitka Slegerova, Miroslav Hajek, Ivan Rehor, Frantisek Sedlak, Jan Stursa,
  Martin Hruby, and Petr Cigler.
\newblock Designing the nanobiointerface of fluorescent nanodiamonds: highly
  selective targeting of glioma cancer cells.
\newblock {\em Nanoscale}, 7(2):415--420, 2015.

\bibitem{song2014statistical}
Xuerui Song, Jian Zhang, Fupan Feng, Junfeng Wang, Wenlong Zhang, Liren Lou,
  Wei Zhu, and Guanzhong Wang.
\newblock A statistical correlation investigation for the role of surface spins
  to the spin relaxation of nitrogen vacancy centers.
\newblock {\em AIP Advances}, 4(4):047103, 2014.

\bibitem{sorkin2002signal}
Alexander Sorkin and Mark von Zastrow.
\newblock Signal transduction and endocytosis: close encounters of many kinds.
\newblock {\em Nature Reviews Molecular Cell Biology}, 3(8):600, 2002.

\bibitem{sotoma2016selective}
Shingo Sotoma, Jun Iimura, Ryuji Igarashi, Koichiro~M Hirosawa, Hidenori
  Ohnishi, Shin Mizukami, Kazuya Kikuchi, Takahiro~K Fujiwara, Masahiro
  Shirakawa, and Hidehito Tochio.
\newblock Selective labeling of proteins on living cell membranes using
  fluorescent nanodiamond probes.
\newblock {\em Nanomaterials}, 6(4):56, 2016.

\bibitem{stacey2018evidence}
Alastair Stacey, Nikolai Dontschuk, Jyh-Pin Chou, David~A Broadway, Alex~K
  Schenk, Michael~J Sear, Jean-Philippe Tetienne, Alon Hoffman, Steven Prawer,
  Chris~I Pakes, Anton Tadich, Nathalie~P. de~Leon, Adam Gali, and Lloyd C.~L.
  Hollenberg.
\newblock Evidence for primal sp2 defects at the diamond surface: candidates
  for electron trapping and noise sources.
\newblock {\em Advanced Materials Interfaces}, page 1801449, 2018.

\bibitem{stehlik2015size}
Stepan Stehlik, Marian Varga, Martin Ledinsky, Vit Jirasek, Anna Artemenko,
  Halyna Kozak, Lukas Ondic, Viera Skakalova, Giacomo Argentero, Timothy
  Pennycook, et~al.
\newblock Size and purity control of hpht nanodiamonds down to 1 nm.
\newblock {\em The Journal of Physical Chemistry C}, 119(49):27708--27720,
  2015.

\bibitem{tanimoto2016detection}
Ryuichi Tanimoto, Takumi Hiraiwa, Yuichiro Nakai, Yutaka Shindo, Kotaro Oka,
  Noriko Hiroi, and Akira Funahashi.
\newblock Detection of temperature difference in neuronal cells.
\newblock {\em Scientific Reports}, 6:22071, 2016.

\bibitem{tisler2009fluorescence}
Julia Tisler, Gopalakrishnan Balasubramanian, Boris Naydenov, Roman Kolesov,
  Bernhard Grotz, Rolf Reuter, Jean-Paul Boudou, Patrick~A Curmi, Mohamed
  Sennour, Alain Thorel, Michael B\"{o}rsch, Kurt Aulenbacher, Rainer Erdmann,
  Philip~R Hemmer, Fedor Jelezko, and J\"{o}rg Wrachtrup.
\newblock Fluorescence and spin properties of defects in single digit
  nanodiamonds.
\newblock {\em ACS Nano}, 3(7):1959--1965, 2009.

\bibitem{tsou2015local}
Chieh-Jui Tsou, Chih-Hao Hsia, Jia-Yin Chu, Yann Hung, Yi-Ping Chen, Fan-Ching
  Chien, Keng~C Chou, Peilin Chen, and Chung-Yuan Mou.
\newblock Local ph tracking in living cells.
\newblock {\em Nanoscale}, 7(9):4217--4225, 2015.

\bibitem{tsuji2017difference}
Toshikazu Tsuji, Kumiko Ikado, Hideki Koizumi, Seiichi Uchiyama, and Kazuaki
  Kajimoto.
\newblock Difference in intracellular temperature rise between matured and
  precursor brown adipocytes in response to uncoupler and $\beta$-adrenergic
  agonist stimuli.
\newblock {\em Scientific Reports}, 7(1):12889, 2017.

\bibitem{turcheniuk2017biomedical}
Kostiantyn Turcheniuk and Vadym Mochalin.
\newblock Biomedical applications of nanodiamond.
\newblock {\em Nanotechnology}, 28:252001, 2017.

\bibitem{tzeng2015time}
Yan-Kai Tzeng, Pei-Chang Tsai, Hsiou-Yuan Liu, Oliver~Y Chen, Hsiang Hsu,
  Fu-Goul Yee, Ming-Shien Chang, and Huan-Cheng Chang.
\newblock Time-resolved luminescence nanothermometry with nitrogen-vacancy
  centers in nanodiamonds.
\newblock {\em Nano Letters}, 15(6):3945--3952, 2015.

\bibitem{wolcott2014surface}
Abraham Wolcott, Theanne Schiros, Matthew~E Trusheim, Edward~H Chen, Dennis
  Nordlund, Rosa~E Diaz, Ophir Gaathon, Dirk Englund, and Jonathan~S Owen.
\newblock Surface structure of aerobically oxidized diamond nanocrystals.
\newblock {\em The Journal of Physical Chemistry C}, 118(46):26695--26702,
  2014.

\bibitem{yamano2017charge}
Hayate Yamano, Sora Kawai, Kanami Kato, Taisuke Kageura, Masafumi Inaba, Takuma
  Okada, Itaru Higashimata, Moriyoshi Haruyama, Takashi Tanii, Keisuke Yamada,
  Shinobu Onoda, Wataru Kada, Osamu Hanaizumi, Tokuyuki Teraji, Junichi Isoya,
  and Hiroshi Kawarada.
\newblock Charge state stabilization of shallow nitrogen vacancy centers in
  diamond by oxygen surface modification.
\newblock {\em Japanese Journal of Applied Physics}, 56(4S):04CK08, 2017.

\bibitem{zhang2011overview}
Yanjie Zhang and Aaron Clapp.
\newblock Overview of stabilizing ligands for biocompatible quantum dot
  nanocrystals.
\newblock {\em Sensors}, 11(12):11036--11055, 2011.

\bibitem{zhao2011chromatographic}
Li~Zhao, Tatsuya Takimoto, Masaaki Ito, Naoko Kitagawa, Takahide Kimura, and
  Naoki Komatsu.
\newblock Chromatographic separation of highly soluble diamond nanoparticles
  prepared by polyglycerol grafting.
\newblock {\em Angewandte Chemie International Edition}, 50(6):1388--1392,
  2011.



\end{thebibliography}

\end{document}